\documentclass{article}
\usepackage{spconf,amsmath,graphicx}
\usepackage{booktabs}
\usepackage{multirow}
\usepackage{bbding}
\usepackage{cite}
\usepackage{bm}
\usepackage{amsfonts,amssymb}
\usepackage{amsmath}
\usepackage{xcolor}
\usepackage[colorlinks = true,
            anchorcolor = black]{hyperref}
\ninept

\newcommand{\tabincell}[2]{\begin{tabular}{@{}#1@{}}#2\end{tabular}}

\title{Connecting Speech Encoder and Large Language Model for ASR}

\name{\parbox{\linewidth}{\centering
Wenyi Yu\textsuperscript{1}, 
Changli Tang\textsuperscript{1}, 
Guangzhi Sun\textsuperscript{1}, 
Xianzhao Chen\textsuperscript{2},\\
Tian Tan\textsuperscript{2},
Wei Li\textsuperscript{2}, 
Lu Lu\textsuperscript{2}, 
Zejun Ma\textsuperscript{2},
Chao Zhang\textsuperscript{1,$\ast$}
\thanks{$\ast$ Corresponding author}
}}

\address{
$^1$ Department of Electronic Engineering, Tsinghua University, $^2$ ByteDance \\ 
\texttt{\small{ywy22@mails.tsinghua.edu.cn; cz277@tsinghua.edu.cn}}
}

\begin{document}
\maketitle
\begin{abstract}
The impressive capability and versatility of large language models (LLMs) have aroused increasing attention in automatic speech recognition (ASR), with several pioneering studies attempting to build integrated ASR models by connecting a speech encoder with an LLM. This paper presents a comparative study of three commonly used structures as connectors, including fully connected layers, multi-head cross-attention, and Q-Former. Speech encoders from the Whisper model series as well as LLMs from the Vicuna model series with different model sizes were studied. Experiments were performed on the commonly used LibriSpeech, Common Voice, and GigaSpeech datasets, where the LLMs with Q-Formers demonstrated consistent and considerable word error rate (WER) reductions over LLMs with other connector structures. Q-Former-based LLMs can generalise well to out-of-domain datasets, where 12\% relative WER reductions over the Whisper baseline ASR model were achieved on the  Eval2000 test set without using any in-domain training data from Switchboard. Moreover, a novel segment-level Q-Former is proposed to enable LLMs to recognise speech segments with a duration exceeding the limitation of the encoders, which results in 17\% relative WER reductions over other connector structures on 90-second-long speech data. 

\end{abstract}

%\blfootnote{Corresponding author$^\ast$}

%
\section{INTRODUCTION}
\label{sec:intro}
Large language models (LLMs) \cite{gpt4,brown2020language,anil2023palm, touvron2023llama,vicuna2023,zeng2021pangu} with rich knowledge and the ability to solve novel and complex tasks have revolutionised the field of natural language processing. 
More recently, significant attention has been drawn to enable LLMs to handle speech inputs \cite{huang2023audiogpt,shen2023hugginggpt,zhang2023speechgpt,rubenstein2023audiopalm,chen2023x,shu2023llasm,wu2023decoder,fathullah2023prompting,li2023prompting,ma2023can,dighe2023leveraging}. In addition to pipeline-based methods in which the LLM serve as a controller to manage a set of functional models \cite{huang2023audiogpt, shen2023hugginggpt}, two categories of approaches have been developed. The first category of approaches discretises speech inputs and embeds the derived speech tokens into a vector space shared with the text tokens, then the LLMs are finetuned to fit into this new token space \cite{zhang2023speechgpt,rubenstein2023audiopalm}. The other category of approaches directly connects a speech encoder with an LLM using a connector that aligns the speech encoders with the LLMs \cite{chen2023x,shu2023llasm,wu2023decoder,fathullah2023prompting,li2023prompting}. This paper focuses on the second category of approaches.

When aligning the speech encoder output and LLM input spaces, the choice of the connector is of vital importance, and should meet the following requirements. First, the connector should be able to retain as much information from the speech inputs as possible, since it determines the amount of information that LLMs can receive from the speech. Second, as the computation and storage costs of LLMs increase considerably when processing long input sequences, the connector should be able to achieve efficient and effective information compression to reduce the lengths of the LLM input sequences.

Focusing on automatic speech recognition (ASR) to show the ability of LLMs to recognise speech, this paper studies different structures of connectors that integrate LLMs with speech encoders. Specifically, an end-to-end ASR system was constructed by connecting a Whisper model encoder \cite{radford2023robust} with a Vicuna LLM \cite{vicuna2023}. Three types of connectors were compared in this paper, including the fully connected layers, multi-head cross-attention \cite{lyu2023macaw} and Q-Former \cite{li2023blip}. To bypass the input length limitation of the pre-trained speech encoders and enable the LLM to process long speech inputs, a novel segment-level Q-Former connector is proposed.
Experiments were conducted based on a training set with $\sim$4,000 hours data, and the LLMs with Q-Former connectors consistently outperform strong Whisper baseline ASR systems on the in-domain datasets, and can achieve over 12\% relative word error rate (WER) reductions on the out-of-domain Eval2000 test set from the Switchboard corpus. The influence of the number of connector output tokens and model size are studied. Moreover, the proposed segment-level Q-Former structure achieved obvious WER reductions on long speech inputs, when compared with other connectors.

%Experiments were performed on around 4,000 hours of training data, where the system with Q-Former connector consistently surpassed Whisper on the in-domain datasets while achieving over 12\% relative word error rate (WER) reductions on the out-of-domain Eval2000 set. Studies were also conducted to analyse the influence of the number of final tokens and model size. Remarkably, the proposed segment-level Q-Former demonstrated its effectiveness on long speech inputs, with obvious WER reductions achieved compared to the aforementioned connectors.

The rest paper is organised as follows. Sec. \ref{sec:related_work} summarises the work related to multimodal LLMs. Secs. \ref{sec:module_connector} and \ref{sec:seg_qformer} introduce the three connectors to compare and the proposed segment-level Q-Former. The experimental setup and results are presented in Secs. \ref{sec:exp_setup} and \ref{sec:exp_results}, followed by conclusions.

\section{RELATED WORK}
\label{sec:related_work}
%\subsection{LLMs with speech inputs}
%\label{subsec:speech_llm}
%Regarding the first category of approaches to extend LLMs for speech, 
%Regarding the first category of approaches to extending LLMs for speech, 
%Regarding the approaches to extending LLMs for speech,
To enable LLMs to perform both speech perception and generation, SpeechGPT \cite{zhang2023speechgpt} and AudioPaLM \cite{rubenstein2023audiopalm} augment the vocabularies of LLaMA \cite{touvron2023llama} and PaLM \cite{anil2023palm} LLMs with discrete speech tokens extracted by HuBERT \cite{hsu2021hubert} and W2v-BERT \cite{chung2021w2v} or USM \cite{zhang2023google} speech encoders respectively. %Surprising speech-to-text/speech translation ability was achieved by AudioPaLM using PaLM as the backbone LLM. 
%Regarding the second category of approaches, 
%
Regarding the approaches to connect multimodal encoders to LLMs, 
X-LLM \cite{chen2023x} interfaced ChatGLM with audio and visual encoders. %LLaSM \cite{shu2023llasm} built a speech LLM with cross-modal conversational abilities. 
\cite{fathullah2023prompting} and \cite{wu2023decoder} connect speech encoders with reduced frame rates to LLMs, and achieve integrated multilingual ASR and speech translation respectively. Moreover, LLMs can also be prompt for domain adaptation \cite{li2023prompting} and uncertainty estimation \cite{dighe2023leveraging} of the ASR results. %The audio reasoning ability of LLMs is studied in \cite{gong2023listen}. MU-LLaMA \cite{liu2023music} showed outstanding music understanding abilities.

Several works studied visual LLMs \cite{alayrac2022flamingo,li2023blip,zhu2023minigpt,dai2023instructblip,zhang2023video,lyu2023macaw,su2023pandagpt,chen2023videollm,videochatgpt}. Following BLIP-2 \cite{li2023blip}, InstructBLIP \cite{dai2023instructblip} and Video-LLaMA \cite{zhang2023video} introduced Q-Former as the module connector. Alternative connectors were also investigated in \cite{alayrac2022flamingo,su2023pandagpt,lyu2023macaw,chen2023videollm}. Regarding audio and music LLMs, the reasoning ability based on audio is studied \cite{gong2023listen}. MU-LLaMA \cite{liu2023music} showed outstanding music understanding abilities.

%Following the first approach, SpeechGPT \cite{zhang2023speechgpt} converted continuous waveforms into tokens with HuBERT \cite{hsu2021hubert}. Then LLaMA \cite{touvron2023llama} model with expanded vocabulary was trained to model both text and audio tokens. AudioPaLM \cite{rubenstein2023audiopalm} followed the same idea and introduced w2v-BERT \cite{chung2021w2v} or USM \cite{zhang2023google} as discrete unit extractors. Surprising speech-to-text/speech translation ability was achieved by AudioPaLM using PaLM as backbone LLM. As for the second approach, X-LLM \cite{chen2023x} proposed X2L interface as intermediate connectors and built an audio-visual LLM. LLaSM \cite{shu2023llasm} built a speech LLM with cross-modal conversational abilities. \cite{wu2023decoder} investigated speech-to-text translation tasks on LLM with a CTC compressor. In \cite{li2023prompting}, the extensive knowledge mastered by LLMs was dug for zero-shot domain adaptation in ASR. \cite{dighe2023leveraging} used LLM to explore the uncertainty in ASR results. Moreover, \cite{fathullah2023prompting} constructed an ASR model by connecting a Conformer \cite{gulati2020conformer} encoder with an LLM.

%\subsection{LLMs with Other Modal Inputs}
%\label{subsec:mm_llm}

% Both SpeechGPT \cite{zhang2023speechgpt} and AudioPaLM \cite{rubenstein2023audiopalm} expanded the vocabulary of LLM with audio tokens while resulted in high computational cost. 

% \subsection{Other multi-modal Large Language Model}

\section{MODULE CONNECTOR}
\label{sec:module_connector}

\begin{figure}[ht]
    \centering
    \includegraphics[width=0.47\textwidth]{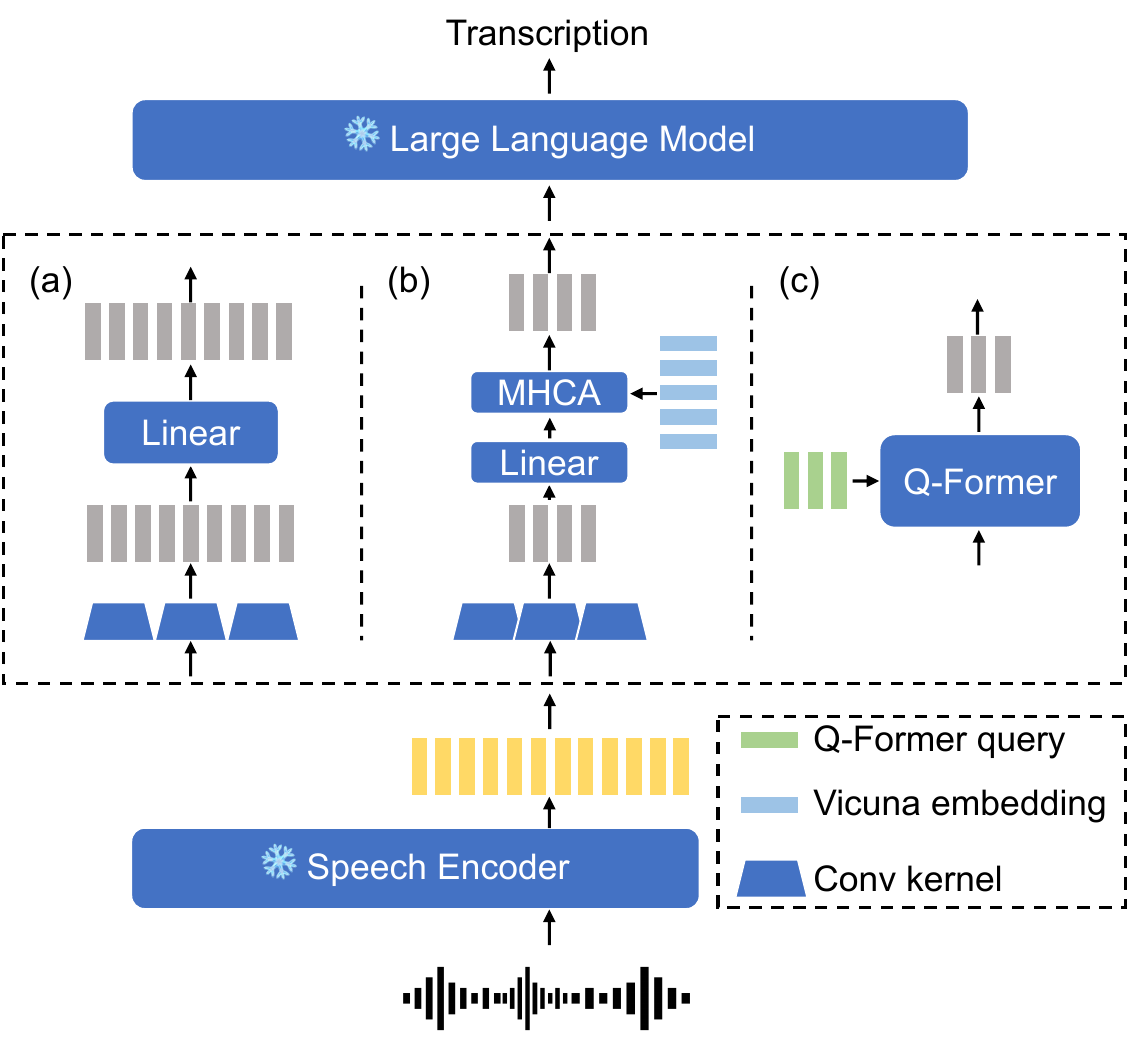}
    \vspace{-0.3cm}
    \caption{Illustration of integrating a speech encoder and an LLM into an ASR system with a module connector of: (a) fully connected layers, (b) multi-head cross-attention, and (c) Q-Former.}
    \label{fig:structure}
\end{figure}

As shown in Fig. \ref{fig:structure}, the proposed ASR model consists of three modules: a frozen speech encoder, a trainable module connector and a frozen LLM. This section introduces three connectors, including fully connected layers, multi-head cross-attention and Q-Former.
\vspace{-0.3cm}
\vspace{-0.3cm}

For clarity, some basic notations are defined as follows: $\mathbf{X}\in \mathbb{R}^{n_x \times d_x}$ denotes the speech features obtained from the speech encoder, and the module connector compresses $\mathbf{X}$ into $\mathbf{T}_\text{speech} \in \mathbb{R}^{n_{t} \times d_{t}}$ which are input to the LLM to produce ASR transcriptions. $\mathbf{H} \in \mathbb{R}^{n_{h} \times d_{h}}$ denotes the hidden states in connectors while $n$ and $d$ are the numbers of vectors and hidden dimensions respectively. %Furthermore, multi-head attention in \cite{vaswani2017attention} is denoted as $\text{MultiHead}(\text{Query},\text{Key},\text{Value})$.

\subsection{Fully connected layers}
\label{subsec:fc}
To compress the length of speech features, $m$ adjacent frames $\mathbf{x}_{i}$, $\mathbf{x}_{i+1}$, ..., $\mathbf{x}_{i+m-1}$ are stacked into $\mathbf{h}_i \in \mathbb{R}^{m\times d_x}$. Then two $\text{Linear}(\cdot)$ layers with $\text{ReLU}(\cdot)$ in between are introduced as follows:
\begin{equation}
    \mathbf{T}_\text{speech}=\text{Linear}(\text{ReLU}(\text{Linear}(\mathbf{H}))),
\end{equation}
where $\mathbf{H}$ consists of $\mathbf{h}_i$ of a batch of samples. 
Actually, the vector stacking operation together with the first linear layer works the same as a 1-dimensional (-d) convolutional layer, $\text{Conv1d}(\cdot)$.

\subsection{Multi-head cross-attention}
\label{subsec:mhsa}
To bridge the gap between the multi-modal encoder output features $\mathbf{X}$ and LLM input textual features $\mathbf{T}_\text{speech}$, a multi-head attention layer \cite{vaswani2017attention} denoted as $\text{MultiHead}(\text{Query},\text{Key},\text{Value})$ is used in the multi-head cross-attention approach to align the two feature spaces \cite{lyu2023macaw}. 
 %$\text{MultiHead}(\cdot)$ layer \cite{vaswani2017attention} between speech features $\mathbf{X}$ and LLM input tokens $\mathbf{T}$. 
First, a $\text{Conv1d}(\cdot)$ layer reduces the length of the speech input by a rate of $s$. Then the hidden states $\mathbf{H}$ are converted to $\mathbf{T}_\text{speech}$ based on the textual embeddings $\mathbf{E}$ using $\text{MultiHead}(\cdot)$. That is,
\begin{align}
    \mathbf{H}&=\text{Linear}(\text{Conv1d}(\mathbf{X}))\\
    \mathbf{T}_\text{speech}&=\text{MultiHead}(\mathbf{H}, \mathbf{E}, \mathbf{E}).
\end{align}

\subsection{Q-Former}
\label{subsec:qformer}

Q-Former \cite{li2023blip} is a Transformer-based module converting variable-length input sequences into fixed-length output query representations. It was initially proposed for visual-text modality alignment, and here is applied to audio-text alignment. In each Q-Former block, trainable query embeddings $\mathbf{Q} \in \mathbb{R}^{n_q \times d_q}$ interact with the input features $\mathbf{X}$ through multi-head self-attention and cross-attention layers, 
%\begin{align}
%\mathbf{T}_s&=\text{MultiHead}(\text{MultiHead}(\mathbf{Q},\mathbf{Q},\mathbf{Q}),\mathbf{X}_s,\mathbf{X}_s).    
%\end{align}
%. The query embeddings can interact with each other through self-attention layers.
Specifically, Q-Former, denoted as $\text{QF}(\mathbf{Q},\mathbf{X})$, in this work consists of two Transformer decoder blocks \cite{vaswani2017attention} with the causal attention masks removed. Here $\mathbf{Q}$ is used as the decoder inputs and $\mathbf{X}$ as the encoder outputs in the standard Transformer.
% \begin{equation}
%     \bm{h}'=\text{LayerNorm}(\text{MultiHead}(\bm{h}, \bm{h}, \bm{h})+\bm{h})
% \end{equation}
% \vspace{-0.5cm}
% \begin{equation}
%     \bm{h}''=\text{LayerNorm}(\text{MultiHead}(\bm{h}', \bm{x}_s, \bm{x}_s)+\bm{h}')
% \end{equation}
% Noted that $\bm{h}=\bm{q}$ for the first layer.

\section{SEGMENT-LEVEL Q-FORMER}
\label{sec:seg_qformer}

\begin{figure}[ht]
    \centering
    \includegraphics[width=0.41\textwidth]{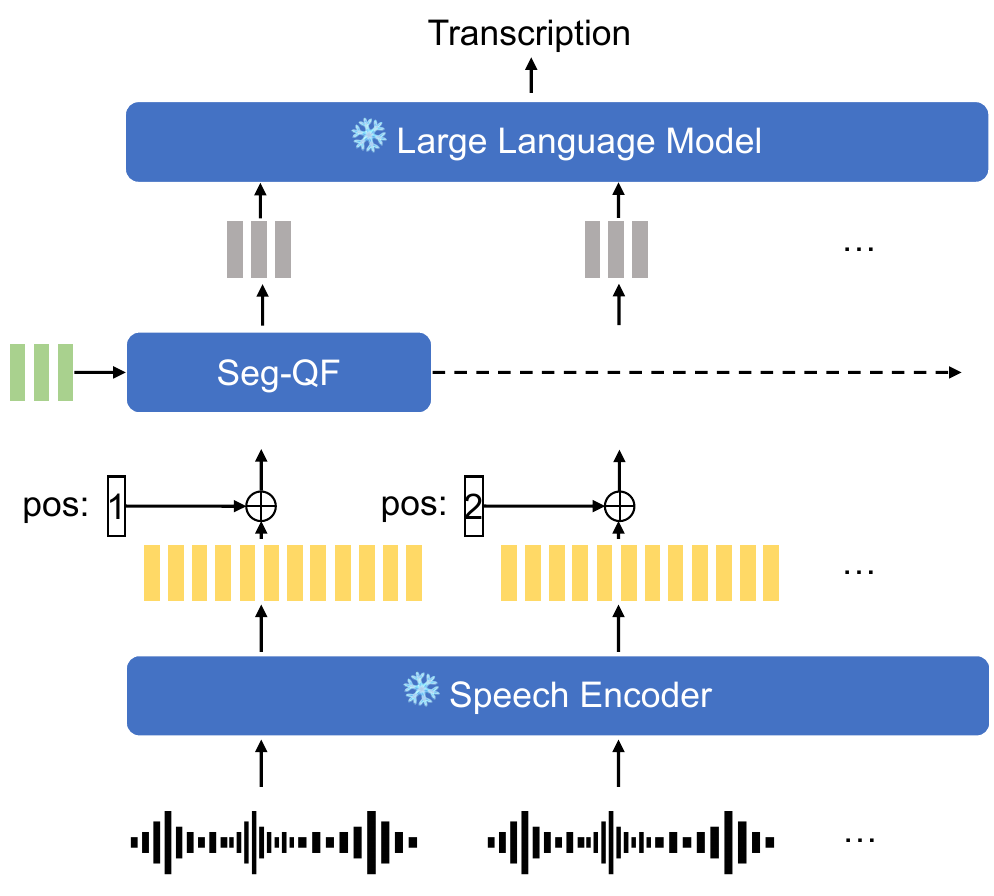}
    \vspace{-0.3cm}
    \caption{The model structure of segment-level Q-Former (seg-QF). The integers in rectangles are segment-level positional encodings.  }
    \label{fig:segQF}
\end{figure}

%Most existing speech encoders have limitations on the duration of processed audio. 
Transformer-based speech encoders can have limitations on the input sequence duration \cite{radford2023robust}.
To enable LLMs to process with longer speech inputs, the whole sequence can be split into several shorter segments to transform by the speech encoder separately. Such segments can be concatenated to reform a single sequence at either the input or output end of Q-Former. In this paper, the structure of segment-level Q-Former (seg-QF)  shown in Fig. \ref{fig:segQF} is proposed, which uses a Q-Former to transform each encoder output segment simultaneously and concatenates their fixed-length output token sequences before feeding into the LLM. Compared to performing the concatenation at the Q-Former input end and producing a fixed number of $n_q$ output tokens, seg-QF allows varying the number of output tokens $N\times n_q$ according to the number of segments $N$, which is more suitable for speech inputs with variable lengths in a wide range.  
Note the trainable query embeddings $\mathbf{Q}$ and Q-Former layers are shared among all the segments, and seg-QF can be initialised with a pre-trained standard Q-Former. 
%However, one of the major differences between Q-Former and other connectors is that Q-Former always generates fixed-length output sequences for variable-length inputs. This is inappropriate when dealing with a variable number of segments as when the input sequence gets longer, fixed-length output queries may not have enough capacity to retain necessary information in the speech. Therefore, a novel segment-level Q-Former (seg-QF) module was proposed as shown in Fig. \ref{fig:segQF}.
\vspace{-0.3cm}
%\vspace{-0.3cm}
%Compared to standard Q-Former, Seg-QF. First, despite that position information is provided by the speech encoder within each segment $\mathbf{S}_i \in \mathbb{R}^{n_s \times d_s}$, $i=1,2,...$, the positional coherency between different segments are not guaranteed. Thus, to inform Q-Former of the position of different segments, segment-level position embeddings $\mathbf{p}_i \in \mathbb{R}^{d_s}$, $i=1,2,...$, are added to the speech features. Second, as the trainable query embeddings $\mathbf{Q}$ and Q-Former layers are shared among all the segments, seg-QF can be initialized with a pre-trained standard Q-Former. Formally, seg-QF is defined as follows:

Despite that relative positions of the frames are provided by the speech encoder within each segment $\mathbf{S}_i\in\mathbb{R}^{n_x\times d_x}$, Seg-QF is not aware of their absolute positions in the whole input sequence.   
To inform Seg-QF with such information, segment-level position embeddings $\mathbf{p}_i \in \mathbb{R}^{d_x}$ are added to $
\mathbf{X}$, as shown in Fig.~\ref{fig:segQF}. Specifically,  
\begin{align}
    \mathbf{T}_\text{speech}&=[\text{QF}(\mathbf{Q},\mathbf{S}_i\oplus \mathbf{p}_i)]^{N}_{i=1},
\end{align}
%\begin{align}
%    \mathbf{T}_\text{speech}&=[\mathbf{T}_{i}]^{N}_{i=1}=[\text{QF}(\mathbf{Q},\mathbf{S}_i\oplus \mathbf{p}_i)]^{N}_{i=1},
%\end{align}
where $\mathbf{S}_i \oplus \mathbf{q}_i$ means adding $\mathbf{q}_i$ to each row of $\mathbf{S}_i$.

\section{EXPERIMENTAL SETUP}
\label{sec:exp_setup}

\subsection{Data specifications}
\label{subsec:training_data}
Experiments were conducted on three setups. Specifically, models in Sections \ref{subsec:rca}-\ref{subsec:model_size} were trained on LibriSpeech \cite{panayotov2015librispeech} 960h dataset. Models in Section \ref{subsec:large_scale_exp} were trained on around 4,000 hours of data including LibriSpeech 960h, Common Voice 7.0 (English) \cite{ardila2020common} and GigaSpeech \cite{GigaSpeech2021} subset M. Models in Section \ref{subsec:exp_segQF} were finetuned on LibriSpeech train-clean-100 subset. While the test sets of the aforementioned data were used for in-domain evaluation, the Eval2000 set was used for out-of-domain evaluation.

\subsection{Model specifications}
\label{subsec:exp_models}

The Vicuna LLMs \cite{vicuna2023}  and the Speech encoders from the Whisper models were used as the decoders and encoders \cite{radford2023robust}, and were both frozen in training. In Section \ref{subsec:model_size}, speech encoders with different sizes were compared, including those from Whisper base, medium and large-v2. 
LLMs including Vicuna 7B and Vicuna 13B were compared. Based on Table~\ref{tab:2}, the best-performing setup with Whisper large-v2 encoder and Vicuna 13B were used in other sections.

%Noted that both the speech encoder and LLM were frozen during training.

%\subsection{Training specifications}
%\label{subsec:training_details}
All the models in Sections \ref{subsec:rca}--\ref{subsec:large_scale_exp} were trained for 90k steps with a batch size of 24 with NVidia A100 80GB GPUs. %In Sections \ref{subsec:rca}--\ref{subsec:model_size}, models were trained on 4$\times$A100 GPUs while for large-scale experiments in Sec. \ref{subsec:large_scale_exp}, models were trained on 32$\times$A100 GPUs. 
Models in Section \ref{subsec:exp_segQF} were initialised with pre-trained Q-Former models or fully connected layers and trained for 10k steps with a batch size of 8. Sinusoid encoding matrix \cite{vaswani2017attention} was used as the segment-level position embeddings. Checkpoints with the highest validation set accuracy obtained with teacher forcing were selected as the final models.

%on the validation set of all the experiments were selected to report the final results. 
%Because Whisper pads/truncates audio samples into 30 seconds by default, if the duration of the input speech is less than 30 seconds, the last few embeddings produced by the Whisper encoder mainly contain padding information. This causes Q-Former models to have \textit{early truncating} problem which results in many deletion errors. Similar to \cite{lin2022random}, a training strategy was proposed that for each utterance in a mini-batch, some other utterances are sampled randomly from the whole training set one by one until the total duration of the concatenated sample is longer than $T$ except that a new utterance makes the total duration longer than 30s. Noted that $T \sim \mathcal{U}[0, 30]$.

% Two scales of experiments were performed in this work. To compare the performance of different connectors and analyse the influence of several key factors, LibriSpeech \cite{panayotov2015librispeech} 960 hours dataset was used for training and word error rates (WER) on LibriSpeech test clean and test other were reported. Moreover, for the large scale experiment, Common Voice 7.0 (EN) \cite{ardila2020common} and GigaSpeech \cite{GigaSpeech2021} subset M were also included, which culminated in about 4,000 hours of training data. Then the performance on the three in-domain datasets and the other two out-domain datasets -- CallHome and Switchboard -- were evaluated to justify the proposed method.

\section{EXPERIMENTAL RESULTS}
\label{sec:exp_results}

\subsection{Random audio concatenation in training}
\label{subsec:rca}
% Because Whisper pads/truncates audio samples into 30 seconds by default, if the duration of the input speech is less than 30s, the last few embeddings produced by the Whisper encoder will contain meaningless information. Although the impact of this problem is negligible in Whisper, experiments showed that Q-Former models trained in this manner often have \textit{early truncating} problem which results in many words deletion. Similar to \cite{lin2022random}, a training strategy was proposed that for each utterance in a batch some other utterances are sampled randomly from the whole training set one by one until the total duration of the concatenated sample is longer than $T$ except that a new utterance makes the total duration longer than 30s. Noted that $T \sim \mathcal{U}[0, 30]$.

%The Whisper encoder is built to have an input duration limitation of 30 seconds, and zero 

The Whisper encoder is built to have a fixed input window of 30 seconds, and zero vectors are padded to the input to increase its sequence length to match the window size. When connecting the Whisper encoder to LLMs using Q-Former and this padding strategy, high deletion errors are often produced for long inputs since the attention mechanisms of Q-Former are trained to ignore the ending elements in the sequence, which are often padded zeros in the training samples.  
%This causes Q-Former models to have \textit{early truncating} problem which results in many deletion errors.
To resolve this issue, a random concatenation strategy is applied, which is similar to that used in  
\cite{lin2022random}. %For each input utterance in a training mini-batch, some other utterances from the whole training set are randomly sampled and concatenated with this input utterance. The concatenation is performed in a one-by-one fashion until the total duration of the concatenated sample is longer than $T$ seconds, where $T$ follows a uniform distribution between 0 and 30 seconds.
Each input utterance in a training mini-batch is concatenated with a number of utterances randomly selected from the whole training set. %The random concatenation stops when the total length of a concatenated utterance exceeds a pre-set upper limit of $T$ seconds. 
The random concatenation continues as long as the total length of the concatenated utterance does not exceed a pre-set upper limit of $T$ seconds. 
The values of $T$ for the training samples are set to follow a uniform distribution of $\mathcal{U}[0, 30]$ seconds.
%, some other utterances from the whole training set are randomly sampled and concatenated with this input utterance. The concatenation is performed in a one-by-one fashion until the total duration of the concatenated sample is longer than $T$ seconds, where $T$ follows a uniformed distribution between 0 and 30 seconds.
%Noted that $T \sim \mathcal{U}[0, 30]$.
%Noted that $T$ follows a uniformed distribution between 0 and 30 seconds. 
%The effectness of the proposed random concatenation of audios is shown in Table \ref{tab:1}. This experiment on Q-Former models which generated 80 tokens as inputs for Vicuna showed that the proposed training strategy can significantly improve the performance. Therefore, in the following experiments, this strategy was used by default.
From the results in Table \ref{tab:1}, it is demonstrated the strategy can considerably reduce deletion errors and improve the WERs, and therefore it is always used in the rest of the experiments.
\begin{table}[ht]
%\footnotesize
    \centering
    \begin{tabular}{c|cccc}
    \toprule
        \multirow{3}{*}{\textbf{\tabincell{c}{Random\\Concat.}}}
        %\textbf{\tabincell{2}{Random\\concat.}} 
        
        & \multicolumn{4}{c}{\textbf{LibriSpeech}} \\
        & \multicolumn{2}{c}{test-clean} & \multicolumn{2}{c}{test-other} \\
        & \%WER & \%Del & \%WER & \%Del \\
    \midrule
        \XSolidBrush & 3.72 &1.38  & 6.47 &1.32 \\
        \Checkmark & \textbf{2.28} & \textbf{0.30} & \textbf{5.20} & \textbf{0.50} \\
    \bottomrule
    \end{tabular}
    \caption{Results with or without using the random concatenation training strategy. Both \%WERs and deletion error rates (\%Del) are shown. The ASR is a Vicuna 13B LLM with a Q-Former connector. }
    \label{tab:1}
\end{table}

\vspace{-0.5cm}
\subsection{Comparisons with different connectors}
\label{subsec:diff_connectors}
A desirable connector should be able to extract all useful information from the input without causing obvious increases in computation and storage. 
%The first ability can be reflected by the final performance of the model and the second property means that the connector cannot generate too many tokens while the total number of trainable parameters is negligible compared to the entire model.
%The first ability can be reflected by the final performance of the model and the second property means that the connector cannot generate too many tokens while the total number of trainable parameters is negligible compared to the entire model.
Regarding ASR, WERs can be used as an indicator to reflect the quality of the information extracted by the connector. %Regarding computation and storage costs, the input sequence length is 
Besides having a reasonable amount of model parameters, the connector is also expected to produce a reduced number of output tokens when required since the number of LLM input tokens is a key factor influencing both LLM computation cost and memory usage.
\begin{table}[t]
%\footnotesize
    \centering
    \begin{tabular}{ccc|cc}
    \toprule
        \multirow{2}*{\textbf{Model}} & \multirow{2}*{\textbf{\#Tokens}} & \multirow{2}*{\textbf{\#Params}} & \multicolumn{2}{c}{\textbf{LibriSpeech}} \\
        %\cline{4-5}
        ~ & ~ & ~ & test-clean & test-other \\
    \midrule
        FC & 75 & 24.6M & 3.00 & 6.70 \\
        FC & 300 & 23.6M & 2.29 & 5.44 \\
        CA & 75 & 133.4M & 3.22 & 7.54 \\
        QF & 60 & 24.5M & 2.33 & 5.43 \\
        QF & 80 & 24.5M & \textbf{2.28} & \textbf{5.20} \\
    \bottomrule
    \end{tabular}
    \caption{\%WERs of LLMs with different connectors. FC, CA, and QF refer to fully connected layers, multi-head cross-attention, and Q-Former connectors respectively.  \#Tokens and \#Params are the numbers of output tokens and model parameters of the connectors.}
    \label{tab:2}
    \vspace{-0.5cm}
\end{table}

In Table \ref{tab:2}, three connectors, including fully connected layers, multi-head cross-attention and Q-Former, are compared. Q-Former results in the lowest WERs on both test sets by producing only 80 output tokens. Although fully connected layers with 300 output tokens (with $m=5$ in Sec.~\ref{subsec:fc})  can achieve similar WERs to Q-Former, it requires much more calculations and memory usage. The WERs produced by the fully connected layers connector with 75 output tokens (with $m=20$) are obviously worse. 
%huge performance degradation can be observed when the number of output tokens is limited to 75. 
The multi-head cross-attention connector \cite{lyu2023macaw} has $\sim$133.4 million (M) model parameters, which are $\sim$6 times more than the others, but still produced the worst WERs. As a result, Q-Former is used in the rest of the study.

\subsection{Trainable queries of Q-Former}
\label{subsec:n_tokens}

Recapping Section \ref{subsec:qformer}, the number of trainable queries used in the Q-Former determines the number of its output tokens. 
%the number of output tokens generated by Q-Former depends on the number of query embeddings. 
This section compares Q-Formers with different numbers of queries. Table \ref{tab:3} shows that increasing the number of queries to 80 can considerably reduce WERs, which implies that using $\sim$80 tokens can retain sufficient information for ASR for inputs with less than 30 seconds (similar WER changes are found when only considering the utterances whose durations are close to 30 seconds), which reveals the strong information compression ability of Q-Former.

% As introduced in Sec. \ref{subsec:qformer}, the number of the tokens generated by Q-Former depends on the number of query embeddings which was analysed in this section. As shown in Tab. \ref{tab:3}, increasing the number of query embeddings can significantly improve the performance of the model until 80. It implies that for speech less than 30 seconds around 80 queries can retain enough information for ASR, which demonstrates the strong information compressing capability of Q-Former.

\begin{table}[ht]
%\footnotesize
    \centering
    \begin{tabular}{c|cc}
    \toprule
        \multirow{2}*{\textbf{\#Queries}} & \multicolumn{2}{c}{\textbf{LibriSpeech}}  \\
        %\cline{2-3}
        ~ & test-clean & test-other \\
    \midrule
        40 & 2.43 & 5.72 \\
        60 & 2.33 & 5.43 \\
        80 & 2.28 & \textbf{5.20} \\
        160 & 2.26 & 5.29 \\
        300 & \textbf{2.19} & 5.25 \\
    \bottomrule
    \end{tabular}
    \caption{Changes in \%WERs by varying the number of trainable queries (equals to the number of output tokens) in Q-Former.}
    \label{tab:3}
\end{table}

\subsection{The sizes of LLMs and speech encoders}
\label{subsec:model_size}
In this section, models with LLMs and speech encoders of different sizes are compared in this section. Q-Former connectors with 80 trainable queries are used for all models. From the results Table~\ref{tab:4}, increasing the sizes of both speech encoder and LLM can result in lower WERs, which is in line with expectations. 
%Tab. \ref{tab:4} tells that better performance can be achieved by using larger and stronger speech encoders and LLMs which is in line with expectations. 
Doubling the size of the speech encoder (Whisper medium to Whisper large) reduces the WERs more obviously than doubling the size of LLMs, which indicates that a stronger speech encoder is more important for ASR, which does not require much content understanding ability. 

\begin{table}[ht]
%\footnotesize
    \centering
    \begin{tabular}{cc|cc}
    \toprule
        \multicolumn{2}{c|}{\textbf{Model scale}} & \multicolumn{2}{c}{\textbf{LibriSpeech}} \\
        %\cline{1-4}
        \textbf{Encoder} & \textbf{LLM} & test-clean & test-other \\
    \midrule
        Whisper base & Vicuna 13B & 3.48 & 9.83 \\
        Whisper medium & Vicuna 13B & 2.35 & 5.66 \\
        Whisper large-v2 & Vicuna 13B & \textbf{2.28} & \textbf{5.20} \\
    %\midrule
        Whisper large-v2 & Vicuna 7B & 2.30 & 5.48 \\
    \bottomrule
    \end{tabular}
    \caption{\%WERs with different speech encoders and LLMs.}
    \label{tab:4}
    \vspace{-0.5cm}
\end{table}

% \begin{table}[ht]
% \tiny
%     \centering
%     \begin{tabular}{ccccccc|cc}
%     \toprule
%         \multirow{2}*{\textbf{Model}} & \multicolumn{2}{c}{\textbf{CommonVoice 7.0}} & \multicolumn{2}{c}{\textbf{LibriSpeech}} & \multicolumn{2}{c|}{\textbf{GigaSpeech}} & \textbf{CallHone} & \textbf{Switchboard} \\
%         \cline{2-9}
%         ~ & \textbf{dev} & \textbf{test} & \textbf{testclean} & \textbf{testother} & \textbf{dev} & \textbf{test} & \textbf{test} & \textbf{test} \\
%     \midrule
%         Whisper & 7.9 & 9.8 & 2.5 & 5.2 & 10.3 & 10.0 & 18.9 & 15.8 \\
%         Whisper\dag &  &  & 2.7 & 5.2 &  &  & 17.6 & 13.8 \\
%          & \textbf{7.1} & \textbf{8.2} & \textbf{2.1} & \textbf{5.0} & \textbf{9.0} & \textbf{9.2} & \textbf{15.5} & \textbf{12.1} \\
%     \bottomrule
%     \end{tabular}
%     \caption{Caption}
%     \label{tab:my_label}
% \end{table}

\subsection{Results with large-scale training set}
\label{subsec:large_scale_exp}
This section verifies the performance of the best-performing model configuration by training on a large-scale dataset with 4,000 hours of speech. A Vicuna 13B LLM is connected with a Whisper large-v2 encoder using a Q-Former connector with 80 trainable queries. The results in Table \ref{tab:5} show that the proposed model outperformed Whisper large-v2 by an average of $\sim$11\% relative WER reduction on the three in-domain test sets. It also generalises well to out-of-domain data by achieving a $\sim$12\% relative WER reduction over Whisper large-v2 on the Eval2000 test sets. 
%, which justifies the proposed ASR framework.
These results verify the superior performance of the proposed model in ASR. 

\begin{table}[ht]
%\footnotesize
    \centering
    \begin{tabular}{l|cc}
    \toprule
        \multirow{2}{*}{\textbf{Test set}} & \multicolumn{2}{c}{\textbf{Model}} \\
        %\cline{3-4}
     ~ & Whisper large-v2 & ours  \\
    \midrule
         LibriSpeech test-clean & 2.5 (2.7) & \textbf{2.1} \\
        LibriSpeech test-other & 5.2 (5.2) & \textbf{5.0} \\
    %\midrule
        Common Voice 7.0 dev & 7.9 (-) & \textbf{7.1} \\
        Common Voice 7.0 test & 9.8 (-) & \textbf{8.2} \\
    %\midrule
        GigaSpeech dev & 10.3 (-) & \textbf{9.0} \\
        GigaSpeech test & 10.0 (-) & \textbf{9.2} \\
    %\midrule
    %\midrule
        CallHome test & 18.9 (17.6) & \textbf{15.5} \\
        Switchboard test & 15.8 (13.8) & \textbf{12.1} \\
    \bottomrule
    \end{tabular}
    \caption{\%WERs of models trained on the large-scale dataset. Numbers in brackets are the official Whisper \%WERs reported in \cite{radford2023robust}.}
    \label{tab:5}
    \vspace{-0.5cm}
\end{table}

\subsection{Results of segment-level Q-Former}
\label{subsec:exp_segQF}
%To evaluate the ability of models to recognise speech exceeding the duration limitation of the speech encoder,
To evaluate the models when recognising speech exceeding the duration limitation of the speech encoder, concatenated long-form test sets were built based on LibriSpeech test-clean and test-other test sets respectively. Utterances of the same chapter were concatenated sequentially to form sequences with a duration limit of 60, 90, or 120 seconds. The models were trained on LibriSpeech 100h subset using random concatenation with $T$ sampled from $\mathcal{U}[0, 90]$. 
%Note that QF represents Seg-QF without the segment-level position embeddings because the standard QF performs much worse.

\begin{figure}
    \begin{minipage}[b]{.48\linewidth}
    \centering
        \centerline{\includegraphics[width=4.0cm]{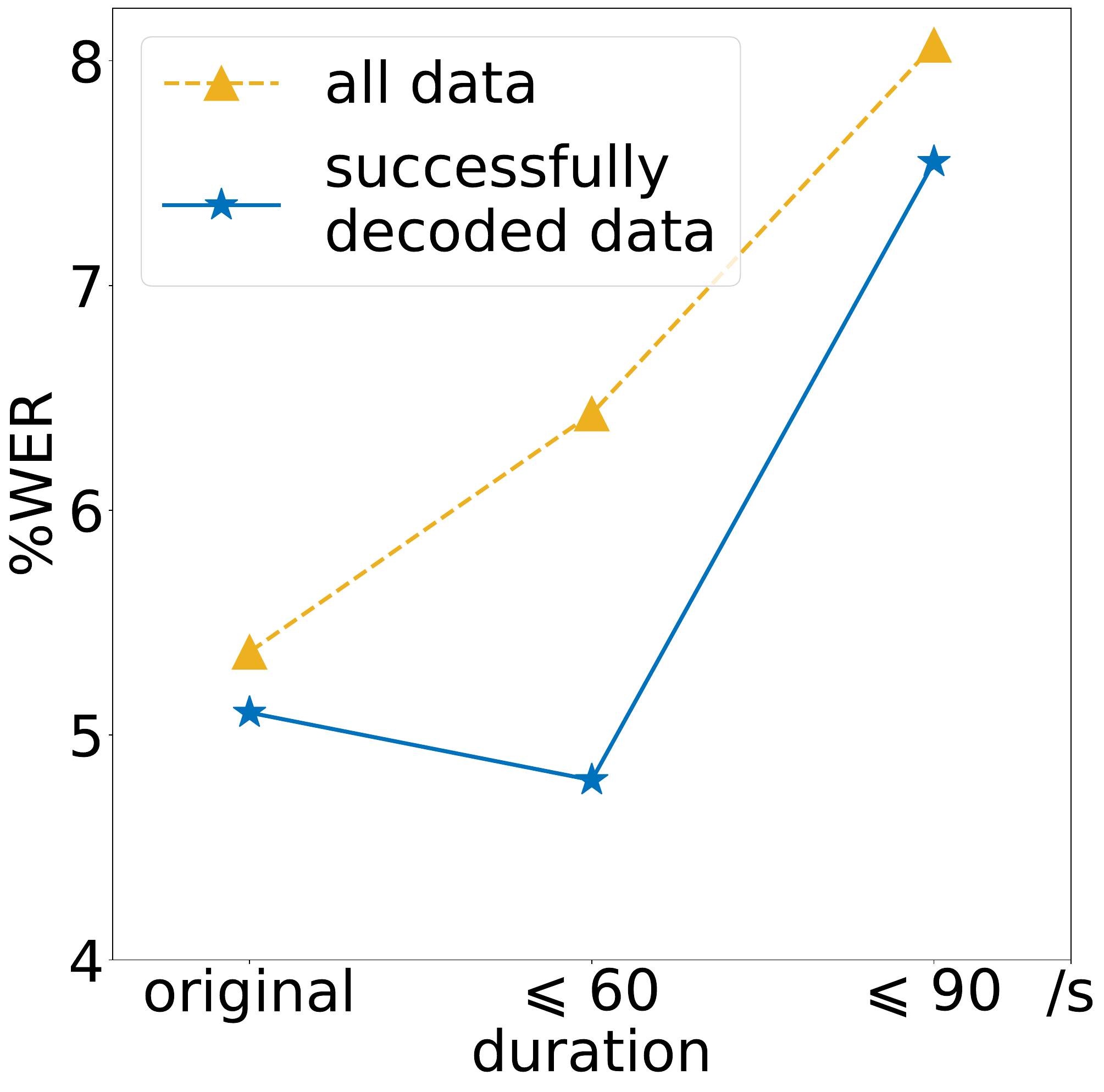}}
    \end{minipage}
    \hfill
    \begin{minipage}[b]{.48\linewidth}
    \centering
        \centerline{\includegraphics[width=4.0cm]{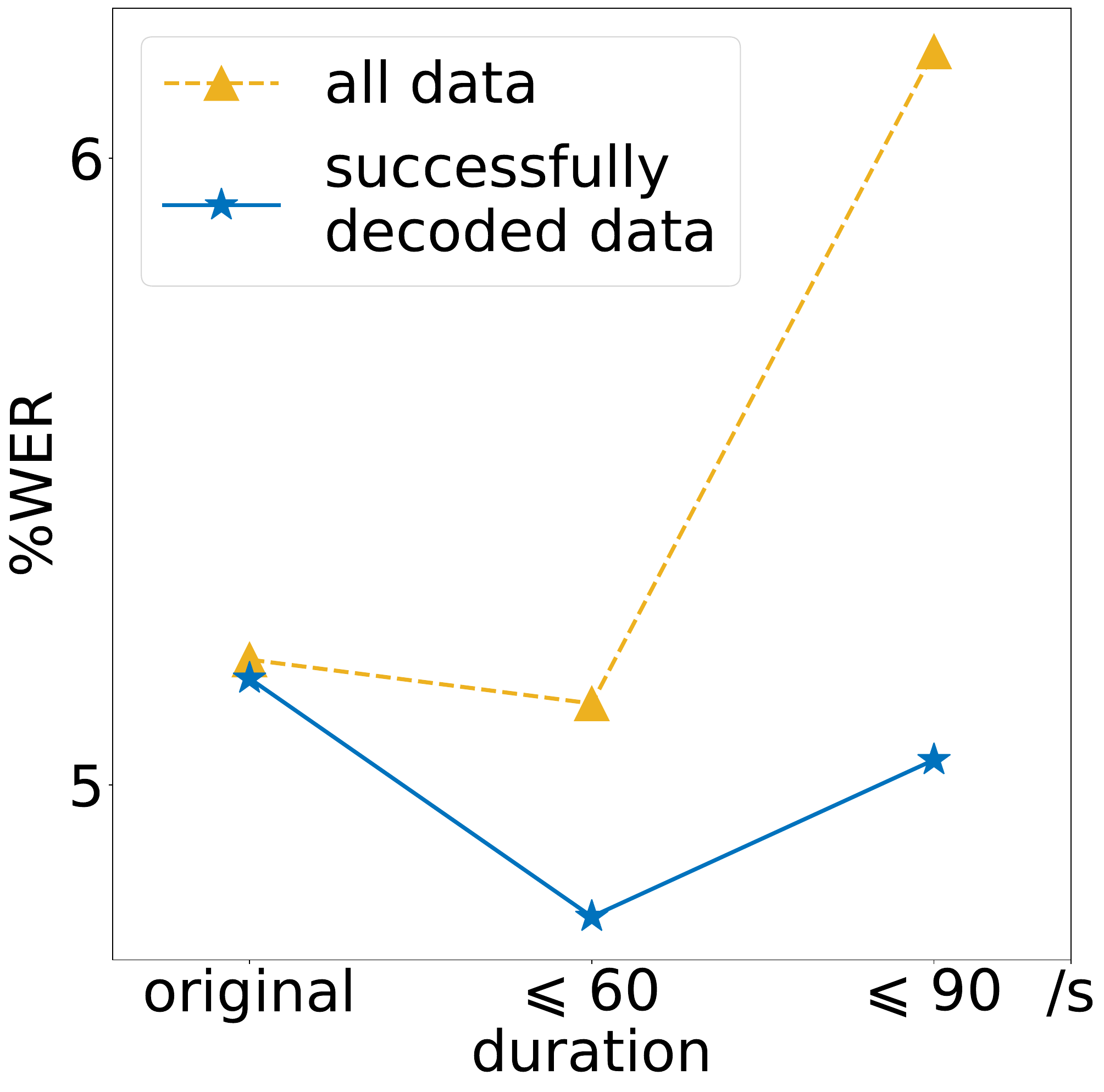}}
    \end{minipage}
    \caption{\%WERs of Seg-QF (\textit{left}) and Seg-QF* (\textit{right}) in Tab. \ref{tab:6} calculated
on all data and successfully decoded data in the test-other set. The x-axis denotes the duration of the concatenated audio.}
    \label{fig:lf_asr}
    \vspace{-0.5cm}
\end{figure}

Different models were compared in Table \ref{tab:6}. Without finetuning on data longer than 30 seconds, QF and FC cannot generalise to recognise longer speech. After finetuning, seg-QF outperformed both QF and FC considerably and showed some generalisation to 120-second-long speech inputs.
From Table \ref{tab:6}, WERs degrade with longer speech inputs. More extreme cases, such as repeated outputs and large chunks of deletions, were observed in case studies when feeding long speech inputs. 
%To better study the samples that have been successfully decoded,
The WERs for all utterances and those successfully decoded in Librispeech test-other are plotted in Fig. \ref{fig:lf_asr} for Seg-QF and Seg-QF*. It shows that the contexts in long speech inputs help ASR to reduce WERs if being decoded successfully.
However, the overall WERs on the full test-other set increased due to more frequent extreme cases. 

\begin{table}[t]
%\footnotesize
\setlength{\tabcolsep}{4pt}
    \centering
    \begin{tabular}{c|cccccc}
    \toprule
        \multirow{2}{*}{\textbf{\tabincell{c}{Concat.\\Test Set}}} & \multicolumn{6}{c}{\textbf{Model}} \\
        %\cline{2-7}
        ~ & FC\dag & QF\dag & FC & QF & Seg-QF & Seg-QF* \\
    \midrule
        test-clean & 3.00 & \textbf{2.28} & 2.89 & 2.32 & 2.35 & \textbf{2.19} \\
        $t\leqslant60$ & 58.18 & 61.99 & 3.43 & 3.83 & \textbf{3.11} & \textbf{2.68} \\
        $t\leqslant90$ & 70.64 & 75.45 & 4.98 & 5.40 & \textbf{3.81} & \textbf{3.24} \\
        $t\leqslant120$ & 75.48 & 76.81 & 19.14 & 26.91 & \textbf{17.95} & \textbf{13.60} \\
    \midrule
        test-other & 6.70 & \textbf{5.20} & 6.68 & 5.40 & 5.37 & \textbf{5.20} \\
        $t\leqslant 60$ & 62.19 & 63.49 & 7.58 & 7.81 & \textbf{6.43} & \textbf{5.13} \\
        $t\leqslant 90$ & 74.27 & 76.03 & 9.03 & 11.78 & \textbf{8.07} & \textbf{6.17} \\
        $t\leqslant 120$ & 78.75 & 76.14 & 24.55 & 32.24 & \textbf{22.25} & \textbf{19.63} \\
    \bottomrule
    \end{tabular}
    \caption{\%WERs on the concatenated test sets. Results based on the Librispeech test clean and test other test sets are shown in Row 1-4 and Row 5-8 respectively, and Row 1 and Row 5 are the results with the original segmentation. Each $t\leqslant T_\text{test}$ test set is obtained by concatenating either test clean or test other utterances based on their orders that appeared in the original audiobooks to up to $T_\text{test}$ seconds.
    Models with \dag \ were not fine-tuned on concatenated training data. Pre-trained checkpoints for Seg-QF* were QF in Table \ref{tab:5},  and were FC in Row 1 and QF in Row 5 of Table \ref{tab:2} for other models.}
    \label{tab:6}
\end{table}

\section{CONCLUSION}
\label{sec:conclusion}
This paper studies to enable LLMs to recognise speech inputs by interfacing with a speech encoder. Three commonly used connectors including fully-connected layers, multi-head cross-attention and Q-Former were compared. The LLMs with Q-Formers demonstrated superior performance over LLMs with other connectors and the Whisper baseline ASR system on all of the in-domain and out-of-domain test sets. Moreover, a novel segment-level Q-Former was proposed to improve the performance with long-form speech inputs whose duration exceeds the limitations of the pre-trained speech encoder. 
Analyses show that although the rich context information in long-form speech inputs can improve ASR accuracy, overly long inputs can also aggravate the hallucination problem of LLMs.

%Further analyses show that the long context helps with ASR while too long audios aggravate the hallucination problem of LLMs.
%At last, a novel segment-level Q-Former was proposed to alleviate the performance degradation when recognising long-form speech inputs whose duration exceeds the limitations of the pre-trained speech encoder. Further analyses show that longer context helps with ASR while too long audios aggravate the hallucination problem of LLMs.

\vfill\pagebreak

\bibliographystyle{IEEEbib}
\bibliography{strings,refs}

\begin{thebibliography}{10}

\bibitem{gpt4}
OpenAI,
\newblock ``{GPT-4} technical report,''
\newblock {\em arXiv:2308.11276}, 2023.

\bibitem{brown2020language}
T.~Brown, B.~Mann, N.~Ryder, et~al.,
\newblock ``Language models are few-shot learners,''
\newblock in {\em Proc. NeurIPS}, 2020.

\bibitem{anil2023palm}
R.~Anil, A.M. Dai, O.~Firat, et~al.,
\newblock ``{PaLM} 2 technical report,''
\newblock {\em arXiv:2305.10403}, 2023.

\bibitem{touvron2023llama}
H.~Touvron, T.~Lavril, et~al.,
\newblock ``{LLaMA}: {O}pen and efficient foundation language models,''
\newblock {\em arXiv:2302.13971}, 2023.

\bibitem{vicuna2023}
W.~Chiang, Z.~Li, Z.~Lin, Y.~Sheng, Z.~Wu, H.~Zhang, L.~Zheng, S.~Zhuang,
  Y.~Zhuang, et~al.,
\newblock ``Vicuna: {A}n open-source chatbot impressing {GPT}-4 with 90\%*
  {ChatGPT} quality,''
\newblock {\em \url{https://vicuna.lmsys.org} (accessed 14 April 2023)}, 2023.

\bibitem{zeng2021pangu}
W.~Zeng, X.~Ren, T.~Su, et~al.,
\newblock ``Pangu-$\alpha$: {L}arge-scale autoregressive pretrained {C}hinese
  language models with auto-parallel computation,''
\newblock {\em arXiv:2104.12369}, 2021.

\bibitem{huang2023audiogpt}
R.~Huang, M.~Li, D.~Yang, et~al.,
\newblock ``{AudioGPT}: {U}nderstanding and generating speech, music, sound,
  and talking head,''
\newblock {\em arXiv:2304.12995}, 2023.

\bibitem{shen2023hugginggpt}
Y.~Shen, K.~Song, X.~Tan, D.~Li, W.~Lu, and Y.~Zhuang,
\newblock ``{HuggingGPT}: {S}olving ai tasks with {chatGPT} and its friends in
  huggingface,''
\newblock {\em arXiv:2303.17580}, 2023.

\bibitem{zhang2023speechgpt}
D.~Zhang, S.~Li, X.~Zhang, J.~Zhan, P.~Wang, Y.~Zhou, and X.~Qiu,
\newblock ``{SpeechGPT}: {E}mpowering large language models with intrinsic
  cross-modal conversational abilities,''
\newblock {\em arXiv:2305.11000}, 2023.

\bibitem{rubenstein2023audiopalm}
P.K. Rubenstein, C.~Asawaroengchai, D.D. Nguyen, et~al.,
\newblock ``{AudioPaLM}: {A} large language model that can speak and listen,''
\newblock {\em arXiv:2306.12925}, 2023.

\bibitem{chen2023x}
F.~Chen, M.~Han, H.~Zhao, Q.~Zhang, J.~Shi, S.~Xu, and B.~Xu,
\newblock ``{X-LLM}: {B}ootstrapping advanced large language models by treating
  multi-modalities as foreign languages,''
\newblock {\em arXiv:2305.04160}, 2023.

\bibitem{shu2023llasm}
Y.~Shu, S.~Dong, G.~Chen, et~al.,
\newblock ``{LLaSM}: {L}arge language and speech model,''
\newblock {\em arXiv:2308.15930}, 2023.

\bibitem{wu2023decoder}
J.~Wu, Y.~Gaur, Z.~Chen, et~al.,
\newblock ``On decoder-only architecture for speech-to-text and large language
  model integration,''
\newblock {\em arXiv:2307.03917}, 2023.

\bibitem{fathullah2023prompting}
Y.~Fathullah, C.~Wu, E.~Lakomkin, J.~Jia, Y.~Shangguan, K.~Li, J.~Guo,
  W.~Xiong, J.~Mahadeokar, O.~Kalinli, et~al.,
\newblock ``Prompting large language models with speech recognition
  abilities,''
\newblock {\em arXiv:2307.11795}, 2023.

\bibitem{li2023prompting}
Y.~Li, Y.~Wu, J.~Li, and S.~Liu,
\newblock ``Prompting large language models for zero-shot domain adaptation in
  speech recognition,''
\newblock {\em arXiv:2306.16007}, 2023.

\bibitem{ma2023can}
R.~Ma, M.~Qian, P.~Manakul, M.~Gales, and K.~Knill,
\newblock ``Can generative large language models perform asr error
  correction?,''
\newblock {\em arXiv:2307.04172}, 2023.

\bibitem{dighe2023leveraging}
P.~Dighe, Y.~Su, S.~Zheng, et~al.,
\newblock ``Leveraging large language models for exploiting asr uncertainty,''
\newblock {\em arXiv:2309.04842}, 2023.

\bibitem{radford2023robust}
A.~Radford, J.~Kim, T.~Xu, G.~Brockman, C.~McLeavey, and I.~Sutskever,
\newblock ``Robust speech recognition via large-scale weak supervision,''
\newblock in {\em Proc. ICML}, Honolulu, 2023.

\bibitem{lyu2023macaw}
C.~Lyu, M.~Wu, L.~Wang, et~al.,
\newblock ``{Macaw-LLM}: {M}ulti-modal language modeling with image, audio,
  video, and text integration,''
\newblock {\em arXiv:2306.09093}, 2023.

\bibitem{li2023blip}
J.~Li, D.~Li, S.~Savarese, and S.~Hoi,
\newblock ``{BLIP}-2: {B}ootstrapping language-image pre-training with frozen
  image encoders and large language models,''
\newblock in {\em Proc. ICML}, Vienna, 2023.

\bibitem{hsu2021hubert}
W.~Hsu, B.~Bolte, Y.~H. Tsai, K.~Lakhotia, R.~Salakhutdinov, and A.~Mohamed,
\newblock ``{HuBERT}: Self-supervised speech representation learning by masked
  prediction of hidden units,''
\newblock {\em IEEE/ACM Transactions on Audio, Speech, and Language
  Processing}, vol. 29, pp. 3451--3460, 2021.

\bibitem{chung2021w2v}
Y.~Chung, Y.~Zhang, W.~Han, C.~Chiu, J.~Qin, R.~Pang, and Y.~Wu,
\newblock ``{W2v-BERT}: Combining contrastive learning and masked language
  modeling for self-supervised speech pre-training,''
\newblock in {\em Proc. ASRU}, Cartagena, 2021.

\bibitem{zhang2023google}
Y.~Zhang, W.~Han, et~al.,
\newblock ``Google {USM}: {S}caling automatic speech recognition beyond 100
  languages,''
\newblock {\em arXiv:2303.01037}, 2023.

\bibitem{alayrac2022flamingo}
J.B. Alayrac, J.~Donahue, P.~Luc, et~al.,
\newblock ``Flamingo: {A} visual language model for few-shot learning,''
\newblock in {\em Proc. NeurIPS}, New Orleans, 2022.

\bibitem{zhu2023minigpt}
D.~Zhu, J.~Chen, X.~Shen, X.~Li, and M.~Elhoseiny,
\newblock ``{MiniGPT}-4: Enhancing vision-language understanding with advanced
  large language models,''
\newblock {\em arXiv:2304.10592}, 2023.

\bibitem{dai2023instructblip}
W.~Dai, J.~Li, D.~Li, A.M.H. Tiong, J.~Zhao, W.~Wang, B.~Li, P.~Fung, and
  S.~Hoi,
\newblock ``{InstructBLIP}: {T}owards general-purpose vision-language models
  with instruction tuning,''
\newblock {\em arXiv:2305.06500}, 2023.

\bibitem{zhang2023video}
H.~Zhang, X.~Li, and L.~Bing,
\newblock ``Video-{LLaMA}: {A}n instruction-tuned audio-visual language model
  for video understanding,''
\newblock {\em arXiv:2306.02858}, 2023.

\bibitem{su2023pandagpt}
Y.~Su, T.~Lan, H.~Li, J.~Xu, Y.~Wang, and D.~Cai,
\newblock ``{PandaGPT}: {O}ne model to instruction-follow them all,''
\newblock {\em arXiv:2305.16355}, 2023.

\bibitem{chen2023videollm}
G.~Chen, Y.D. Zheng, et~al.,
\newblock ``{VideoLLM}: Modeling video sequence with large language models,''
\newblock {\em arXiv:2305.13292}, 2023.

\bibitem{videochatgpt}
M.~Maaz, H.~Rasheed, S.~Khan, and F.S. Khan,
\newblock ``{Video-ChatGPT}: Towards detailed video understanding via large
  vision and language models,''
\newblock {\em arXiv:2306.05424}, 2023.

\bibitem{gong2023listen}
Y.~Gong, H.~Luo, A.H. Liu, L.~Karlinsky, and J.~Glass,
\newblock ``Listen, think, and understand,''
\newblock {\em arXiv:2305.10790}, 2023.

\bibitem{liu2023music}
S.~Liu, A.~Hussain, C.~Sun, and Y.~Shan,
\newblock ``Music understanding {LLaMA}: Advancing text-to-music generation
  with question answering and captioning,''
\newblock {\em arXiv:2308.11276}, 2023.

\bibitem{vaswani2017attention}
A.~Vaswani, N.~Shazeer, N.~Parmar, et~al.,
\newblock ``Attention is all you need,''
\newblock in {\em Proc. NeurIPS}, Long Beach, 2017.

\bibitem{panayotov2015librispeech}
V.~Panayotov, G.~Chen, D.~Povey, and S.~Khudanpur,
\newblock ``Librispeech: {A}n {ASR} corpus based on public domain audio
  books,''
\newblock in {\em Proc. ICASSP}, South Brisbane, 2015.

\bibitem{ardila2020common}
R.~Ardila, M.~Branson, K.~Davis, et~al.,
\newblock ``Common {V}oice: {A} massively-multilingual speech corpus,''
\newblock in {\em Proc. LREC}, Marseille, 2020.

\bibitem{GigaSpeech2021}
G.~Chen, S.~Chai, G.~Wang, et~al.,
\newblock ``{GigaSpeech}: {A}n evolving, multi-domain asr corpus with 10,000
  hours of transcribed audio,''
\newblock in {\em Proc. Interspeech}, Brno, 2021.

\bibitem{lin2022random}
Y.~Lin, T.~Han, H.~Xu, et~al.,
\newblock ``Random utterance concatenation based data augmentation for
  improving short-video speech recognition,''
\newblock in {\em Proc. Interspeech}, Dublin, 2023.

\end{thebibliography}

\end{document}